\documentclass[%
 aip,
 apl,
 amsmath,amssymb,
 reprint,%
]{revtex4-1}

\usepackage{natbib}
\usepackage{graphicx}
\usepackage{dcolumn}
\usepackage{bm}

\usepackage[T1]{fontenc}
\usepackage{mathptmx}
\usepackage{color}
\begin{document}
%
\title{Determination of spin relaxation time and spin diffusion length by oscillation of spin pumping signal}
\date{\today}
\author{Junji Fujimoto}
\email[E-mail address: ]{fujimoto.junji@gmail.com}
\affiliation{Kavli Institute for Theoretical Sciences, University of Chinese Academy of Sciences, Beijing, 100190, China}
\author{Mamoru Matsuo}%
\affiliation{Kavli Institute for Theoretical Sciences, University of Chinese Academy of Sciences, Beijing, 100190, China}
\begin{abstract}
We theoretically investigate a manipulation method of nonequilibrium spin accumulation in the paramagnetic normal metal of a spin pumping system, by using the spin precession motion combined with the spin diffusion transport.
We demonstrate based on the Bloch-Torrey equation that the direction of the nonequilibrium spin accumulation is changed by applying an additional external magnetic field, and consequently, the inverse spin Hall voltage in an adjacent paramagnetic heavy metal changes its sign.
We find that the spin relaxation time and the spin diffusion length are simultaneously determined by changing the magnitude of the external magnetic field and the thickness of the normal metal in a commonly-used spin pumping system.
\end{abstract}
\maketitle
Recent topics in spintronics are mostly concerned with generation and detection of spin.
A variety of methods of the generating spin accumulation and spin current have been investigated in the past few decades, such as the spin pumping effect~\cite{mizukami2002,tserkovnyak2002,tserkovnyak2005}, spin Hall effect~\cite{murakami2003,sinova2004,kato2004,valenzuela2006,kimura2007,sinova2015}, spin Seebeck effect~\cite{uchida2008,uchida2011,xiao2010,adachi2010}, gyromagnetic effect~\cite{matsuo2011,matsuo2013,takahashi2016a,matsuo2017,hirohata2018} and nonlocal spin-valve devices~\cite{Zutic2004}.
The generated spin is detected by using reciprocal phenomena of the generation, such as the inverse spin Hall effect~\cite{saitoh2006,kimura2007,vila2007}, ac spin current detection by spin-transfer torque ferromagnetic resonance~\cite{kobayashi2017}, and also detected optically~\cite{fiederling1999,hirohata2001,puebla2017,stamm2017,hirohata2018}.

The manipulation of generated spin accumulation and spin current is one of the most important issues, which has, however, few attention in spintronics.
The spin accumulation or magnetization can be controlled by spin diffusion transport~\cite{torrey1956} and by spin precession motion~\cite{bloch1946}.
The spin diffusion transport is controlled by changing the material parameters, the thickness of materials, and temperature~\cite{yang2018}.
For the spin precession motion, one changes the direction of the spin by applying the magnetic field or by current-induced spin torques~\cite{Tatara2008}.

For paramagnetic normal metals~(NM), the spin diffusion transport is commonly considered to manipulate the nonequilibrium spin accumulation~\cite{tserkovnyak2005,ando2011,rojas-sanchez2014,yang2018}.
 It is analyzed based on the spin diffusion equation, $\lambda^2 \bm{\nabla}^2 \mu^{\alpha} = \mu^{\alpha}$, where $\mu^{\alpha}$ ($\alpha = x, y, z)$ is the nonequilibrium spin accumulation.
Here, $\lambda$ is the spin diffusion length, which is the only physical parameter in the equation, and the three components of the accumulation are independent of each other. 
On the other hand, the spin precession due to the external magnetic field in NM is not used except in nonlocal spin-valve devices~\cite{Zutic2004}, such as the Hanle measurement~\cite{johnson1985}.

In this letter, we consider the spin diffusion transport as well as spin precession motion, and show that both the spin relaxation time and the spin diffusion length of NM can be determined in spin pumping systems with a contiguous paramagnetic heavy metal~(HM), without any material parameter changed. 
We also show that the spin pumping signal in HM can oscillate due to the spin precession in NM.

Both the spin relaxation time and the spin diffusion length are essential quantities in spintronics.
However, compared to the spin diffusion length, there are fewer techniques to evaluate the spin relaxation time, such as the transmission electron spin resonance technique~\cite{lewis1967} and the Hanle measurement.~\cite{johnson1985}
These methods require a sophisticated experimental setup to carry out.
One of the advantages of our method proposed here is to enable us to measure the spin relaxation time in a commonly-used spin pumping system.  

\begin{figure}[tbph]
\centering
\includegraphics[width=0.9\linewidth,clip]{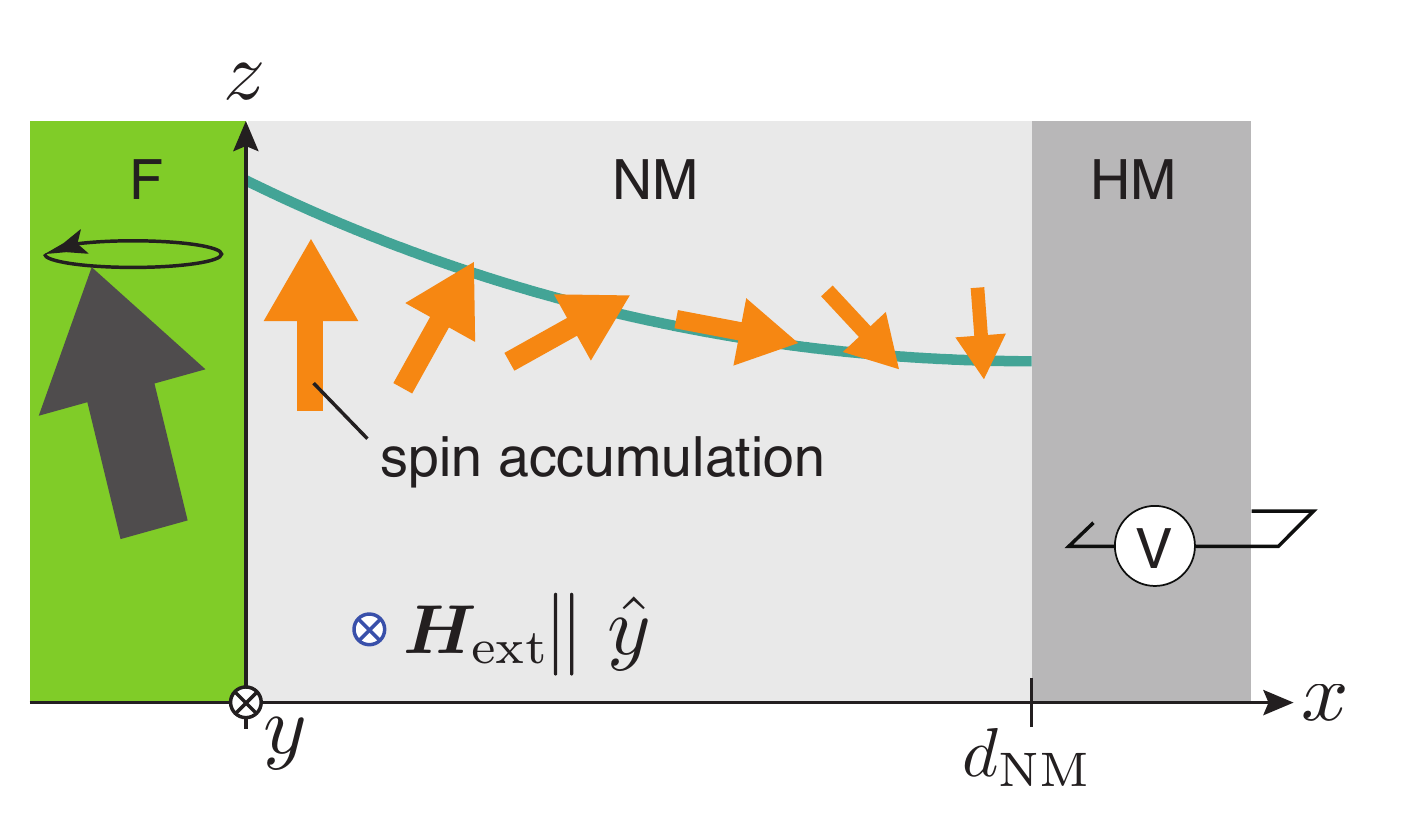}
\caption{\label{fig:1}Schematic description of the concept of this letter.
The magnetization precession in a ferromagnet~(F) induces a nonequilibrium spin accumulation in NM, which diffuses with precession by the external magnetic field.
The direction of the spin accumulation at the other side of NM depends on the strength of the applied magnetic field.
}
\end{figure}

We first consider NM attached to a ferromagnet~(F) without HM, for simplicity.
The configuration is widely used for the spin pumping effect, where the magnetization precession in F due to the ferromagnetic resonance~(FMR) injects spin angular momentum into the NM layer, which causes nonequilibrium spin accumulation in NM.
When the external magnetic field is absent in NM, the spin accumulation simply obeys the spin diffusion equation.
Conversely, we consider the case in the presence of the external magnetic field in NM, where the spin accumulation obeys the Bloch-Torrey~(BT) equation\cite{torrey1956,johnson1985}~\footnote{The original Bloch-Torrey equation can apply to more general cases that the time dependence is treated, the longitudinal relaxation $T_1$ is different from the transverse one $T_2$, and the diffusion constant $D$ can depend on space. We restrict the condition, in which the dynamical scale is about FMR scale, so that we can consider the steady state, the longitudinal and transverse relaxation times are equivalent, $T_1 = T_2 = \tau_{\rm sf}$ in NM, and $D$ is constant for space.}\footnote{The spin relaxation time and the spin diffusion length is connected to each other in the form, $\lambda = \sqrt{D \tau_{\rm sf}}$.},
\begin{align}
\lambda^2 \bm{\nabla}^2 \bm{\mu}
	& = \bm{\mu}
		- \tau_{\rm sf} \gamma \bm{\mu} \times \bm{H}_{\rm ext}
		- \chi \bm{H}_{\rm ext}
\label{eq:BTeq}
.\end{align}
Here, $\gamma$ is the gyromagnetic ratio for electron spin, $\bm{H}_{\rm ext}$ is the external magnetic field, $\tau_{\rm sf}$ is the spin relaxation time, and $\chi$ is the magnetic susceptibility.
For the case $\bm{H}_{\rm ext} = 0$, the BT equation reduces to the spin diffusion equation.
The second term in the right-hand side of Eq.~(\ref{eq:BTeq}) describes the precession of the nonequilibrium spin accumulation, and the third denotes the relaxation of the spin accumulation along the magnetic field.
Note that each component of the accumulation in the BT equation is no longer independent.

For the simple configuration consisting of F and NM (without HM), the nonequilibrium spin accumulation depends only on the distance from the interface of F and NM, $\mu^{\alpha} = \mu^{\alpha} (x)$.
The external magnetic field is applied along the $y$ direction, $\bm{H}_{\rm ext} = H_{\rm ext} \hat{y}$.
We solve the BT equation numerically under the following boundary conditions for the spin current defined by $j^{\alpha}_{s, x} (x) = - (\sigma / e) \mathrm{d} \mu^{\alpha} / \mathrm{d} x$, where $\sigma$ is the conductivity of NM and $e$ is the elementary charge; (i) the spin current at the interface of F and NM takes a certain value $\bm{j}_{s, x} (0) = (0, 0, j_{s0})$, which is corresponding to the flow of the angular momentum injected by the spin pumping effect, and (ii) the spin current is zero at the other boundary of NM, $\bm{j}_x (d_{\rm NM}) = 0$, where $d_{\rm NM}$ is the thickness of NM.

\begin{figure}[tbp]
\centering
\includegraphics[width=0.9\linewidth]{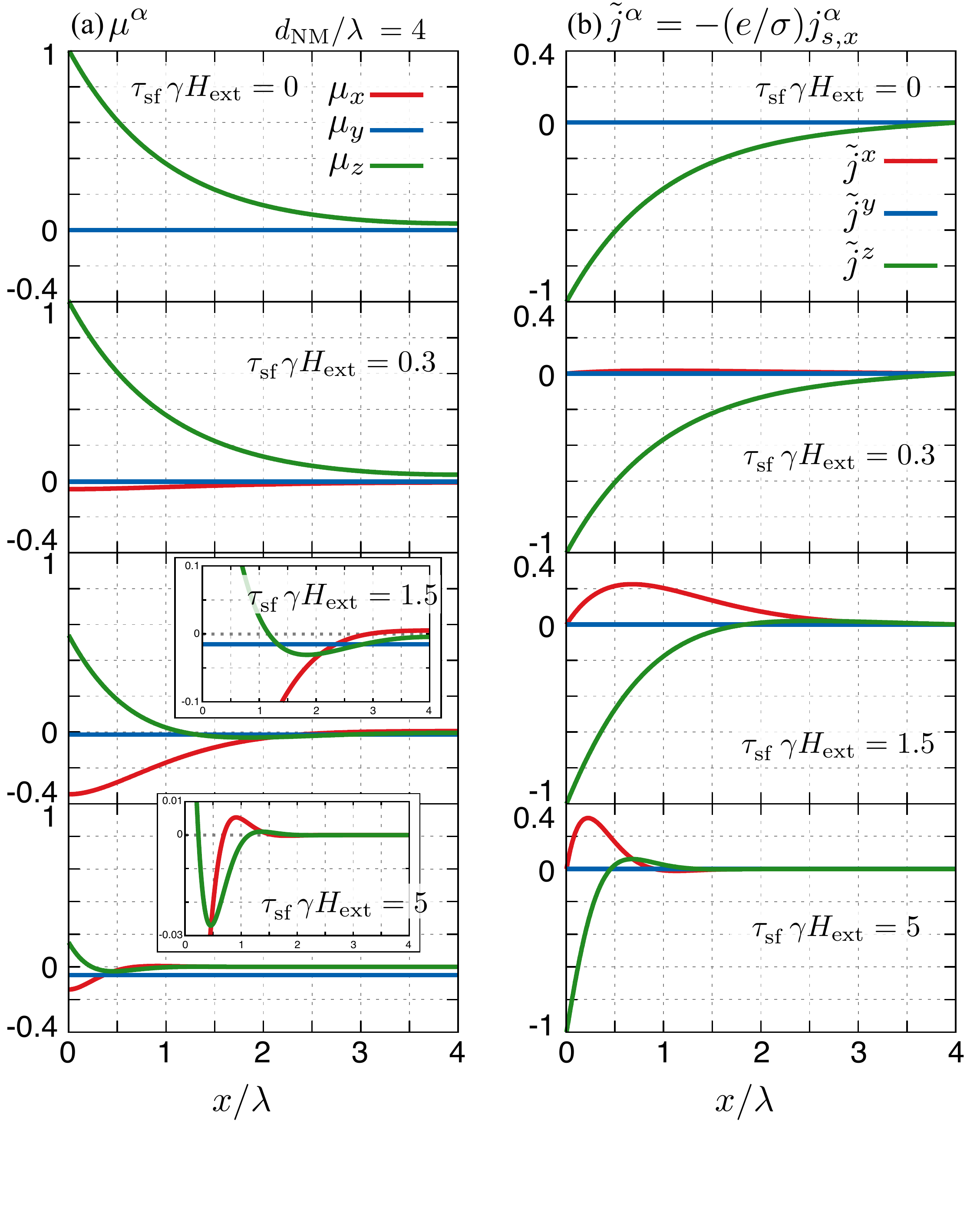}
\caption{\label{fig:2} The distribution of (a) the nonequilibrium spin accumulation and (b) spin current in NM for $d_{\rm NM} / \lambda = 4$.
Depending on the magnitude of the external magnetic field $H_{\rm ext}$, the distribution changes, and both the spin accumulation and spin current take negative values when $\tau_{\rm sf} \gamma H_{\rm ext} \gtrsim 1$.}
\end{figure}
The calculated 
result is shown in Fig.~\ref{fig:2}, in which (a) depicts the distribution of the spin accumulation $\mu^{\alpha}(x)$ and (b) is that of the corresponding spin current given by $j^{\alpha}_{s, x} (x)$.
When the external magnetic field is absent, $H_{\rm ext} = 0$, the only $z$ component of the spin accumulation diffuses, corresponding to the solutions obtained from the spin diffusion equation.
For $\tau_{\rm sf} \gamma H_{\rm ext} \simeq 1$, which is equivalent to $\tau_{\rm sf} \simeq 10 \text{ps}$ for $H_{\rm ext} = 1 \mathrm{T}$, the $x$ component also arises in the same order as $z$ component and diffuses.

One of the crucial points of this letter is that the $z$ component of the nonequilibrium spin accumulation can take the negative value with oscillation, when $\tau_{\rm sf} \gamma H_{\rm ext} \gtrsim 1$.
This indicates that the external magnetic field modifies the transport of the nonequilibrium spin accumulation.
Note that the larger magnitude of the external field is needed for the thinner NM in order to change the sign of the spin accumulation.

Next, we consider the trilayer structure consisting of F, NM, and HM, as shown in Fig.~\ref{fig:1}, to detect the spin accumulation by the inverse spin Hall current of HM.
In this situation, we apply the external magnetic field along the $y$ direction, $\bm{H}_{\rm ext} = H_{\rm ext} \hat{y}$.
We now solve the BT equation for NM and HM, $\lambda_{a}^2 \bm{\nabla}^2 \bm{\mu}_{a} = \bm{\mu}_{a} - \tau_{\rm sf}^{a} \gamma \bm{\mu}_{a} \times \bm{H}_{\rm ext} - \chi \bm{H}_{\rm ext}$, where $a \in \{\mathrm{NM}, \mathrm{HM}\}$, under the boundary conditions, in addition to (i), (ii$^\prime$) the spin current vanishes at the surface of HM, $j^{\alpha}_{s, x} (d_{\rm NM}+d_{\rm HM}) = 0$ with $d_{\rm HM}$ being the thickness of HM, (iii) the spin accumulation is continuous $\mu^{\alpha}_{\rm NM} (d_{\rm NM}) = \mu^{\alpha}_{\rm HM} (d_{\rm NM})$, and (iv) the spin current is also continuous, $
 j^{\alpha}_{s, x} (d_{\rm NM})
	= - \frac{\sigma_{\rm NM}}{e} \frac{\mathrm{d} \mu_{\rm NM}^{\alpha}}{\mathrm{d} x}
	= - \frac{\sigma_{\rm HM}}{e} \frac{\mathrm{d} \mu_{\rm HM}^{\alpha}}{\mathrm{d} x}
$, at the interface between NM and HM.

\begin{figure*}[htbp]
\centering
\includegraphics[width=0.9\linewidth]{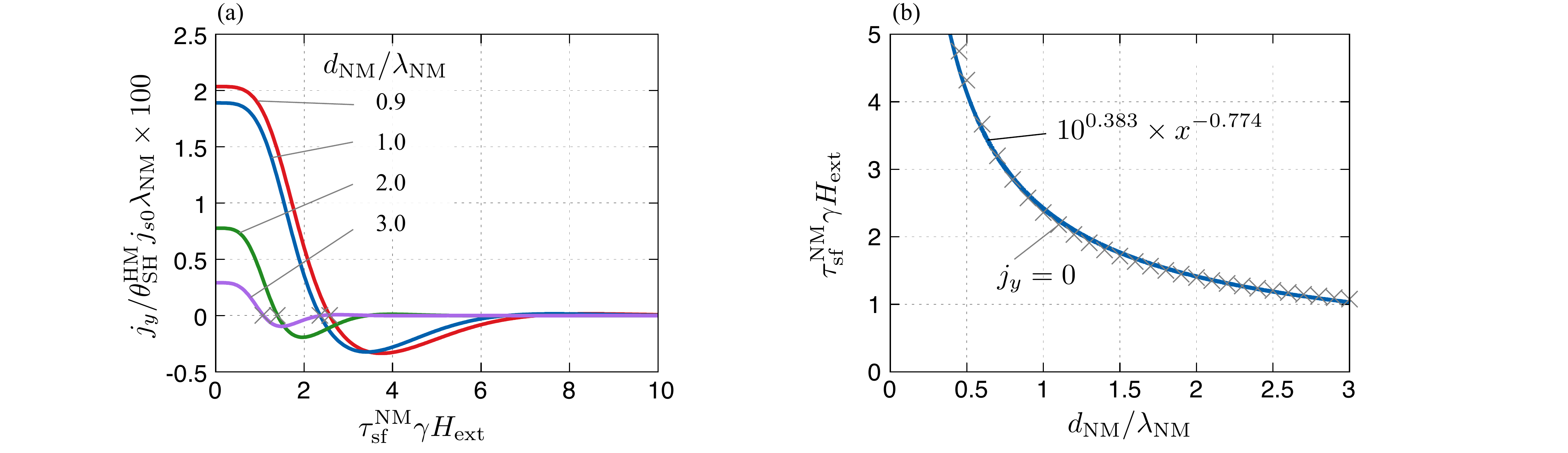}
\caption{\label{fig:3}(a) The inverse spin Hall current $j_y$ as a function of the external magnetic field.
(b) The vanishing point of the current is plotted by the ``$\times$" symbol as a function of the thickness of NM.
The physical values are used for the case of $\mathrm{Al}$ as the NM and $\mathrm{Pt}$ as the HM shown in the main text.
}
\end{figure*}

The calculated distribution of the nonequilibrium spin accumulation of NM, $\mu^{z}_{\rm NM}$, is not changed qualitatively from in the two-layer structure of F and NM, in which the sign inversion of $\mu^{z}_{\rm NM}$ occurs when $\tau_{\rm sf}^{\rm NM} \gamma H_{\rm ext} \gtrsim 1$.
Because this spin accumulation is injected into HM, the electric current by the inverse spin Hall effect,
\begin{align}
j_{y}
	& = \int_{d_{\rm NM}}^{d_{\rm NM}+d_{\rm HM}} \theta_{\rm SH}^{\rm HM}\left( - \frac{\sigma_{\rm HM}}{e} \frac{\mathrm{d} \mu^{z}_{\rm HM}}{\mathrm{d} x} \right) \mathrm{d}x
\notag \\
	& = \theta_{\rm SH}^{\rm HM} \frac{\sigma_{\rm HM}}{e} \left\{ \mu^{z}_{\rm HM} (d_{\rm NM}) - \mu^{z}_{\rm HM} (d_{\rm NM} + d_{\rm HM}) \right\}
,\end{align}
is obtained as shown in Fig.~\ref{fig:3}~(a), and the sign of the current also changes positive to negative.
In the HM layer, the spin accumulation $\mu^{\alpha}_{\rm HM}$ just diffuses since $\tau_{\rm sf}^{\rm HM} \gamma H_{\rm ext} \ll 1$ in the realistic situation since $\tau_{\rm sf}^{\rm HM} \ll \tau_{\rm sf}^{\rm NM}$.

As increasing the external magnetic field, the precession angle of the spin accumulation becomes larger during the diffusion.
For the $z$ component of the spin accumulation, there is a certain magnitude of the external field, where the precession angle is equivalent to $\pi/2$ during the diffusion, in which the inverse spin Hall current vanishes.
 The vanishing inverse spin Hall current corresponds to the zero spin accumulation at $x = d_{\rm NM}$ in the corresponding component, since $|\mu^{\alpha} (d_{\rm NM})| > | \mu^{\alpha} (d_{\rm NM}+d_{\rm HM})|$ unless $\mu^{\alpha} (d_{\rm NM}) = 0$.
This particular magnitude is plotted as a function of the thickness of NM in Fig.~\ref{fig:3}~(b).
 By fitting this curve using the experimental data, 
 we identify the $d_{\rm NM} / \lambda_{\rm NM}$, and $\tau^{\rm NM}_{\rm sf} \gamma H_{\rm ext}$. 
Both the spin relaxation time $\tau_{\rm sf}^{\rm NM}$ and the spin diffusion length $\lambda_{\rm NM}$ can be determined using known parameters, $d_{\rm NM}$, $H_{\rm ext}$ and $\gamma$.
It should be noted that the vanishing point of $j_y$ does not depend on any material parameters of HM but the spin accumulation and spin current injected from NM for the case of $\tau^{\rm HM}_{\rm sf} \ll \tau^{\rm NM}_{\rm sf}$.
It is also noted that the inverse spin Hall current becomes larger as the ratio $\sigma_{\rm NM} / \sigma_{\rm HM}$ is smaller, while the qualitative behavior of the current does not change depending on the ratio.

Finally, we evaluate the order of the inverse spin Hall current in HM.
We consider a case of $\mathrm{Al}$ as the NM and $\mathrm{Pt}$ as the HM with the injected spin current $j_{s0} \sim 10^{7} \, \mathrm{A}\, \mathrm{m}^{-2} $,~\cite{rojas-sanchez2014} and the spin Hall angle being $\theta_{\rm SH}^{\rm HM} = 0.04$.~\cite{ando2011}
Using $\sigma_{\rm NM} = 1.7 \times 10^{7} \,\mathrm{\Omega}^{-1} \mathrm{m}^{-1}$, $\lambda_{\rm NM} = 650 \mathrm{nm}$, and $\tau^{\rm NM}_{\rm sf} = 100 \times 10^{-12}\, \mathrm{s}$,~\cite{jedema2002} $\sigma_{\rm HM} = 1 / (42 \times 10^{-9} \, \mathrm{\Omega} \mathrm{m})$, $\lambda_{\rm HM} = 14 \mathrm{nm}$,~\cite{kurt2002} and $\tau_{\rm sf}^{\rm HM} \sim 1.6 \times 10^{-14} \, \mathrm{s}$,~\cite{freeman2018} we obtain $j_y \sim 0.1 \times 10^{4} \lambda_{\rm NM} \,\mathrm{A}\, \mathrm{m}^{-2} = 6.5 \times 10^{-2} \, \mathrm{A}\, \mathrm{cm}^{-1}$, which could be detectable in an experiment.
Since the gyromagnetic ratio is $\gamma = 1.760 \times 10^{11}\, \mathrm{rad}\, \mathrm{s}^{-1}\, \mathrm{T}^{-1}$, for the factor $\tau^{\rm NM}_{\rm sf} \gamma H_{\rm ext} \sim 5 $, it is necessary to apply the external magnetic field $H_{\rm ext} \sim 0.29 \mathrm{T}$, which is reachable in an experiment.

In Conclusion, we demonstrated that the nonequilibrium spin accumulation is controlled by applying the external magnetic field combined with the diffusion transport.
 Furthermore, we showed the oscillation of the spin pumping signal, i.e., the sign change of the inverse spin Hall current $j_y$ in the adjacent HM by increasing the magnitude of the external field.
From the point of the sign changing, we can determine both the spin relaxation time and the spin diffusion length without any material parameters changed.

\nocite{*}
\bibliography{reference}
\end{document}